\documentclass[prl,twocolumn,epsfig,rotate,showpacs]{revtex4}
\usepackage{epsfig}
\usepackage{graphicx}
\usepackage{amsmath}

\newcommand{\beq}{\begin{eqnarray}}
\newcommand{\eeq}{\end{eqnarray}}
\begin{document}


\title{Towards a Cold-atom Realization of Quantum Maps with Hofstadter's Butterfly
Spectrum}
\author{Jiao Wang$^{1}$ and Jiangbin Gong$^{2,3}$}
\email{phygj@nus.edu.sg} \affiliation{$^{1}$Temasek Laboratories and
Beijing-Hong Kong-Singapore Joint Centre for Nonlinear and Complex
Systems (Singapore), National University of Singapore, 117542,
Singapore
\\ $^{2}$Department of
Physics and Centre of Computational Science and Engineering,
National University of Singapore, 117542, Singapore
\\ $^{3}$ NUS Graduate School for Integrative Sciences and
Engineering, Singapore
 117597, Republic of Singapore}
\date{Oct. 4, 2007}

\begin{abstract}
Quantum systems with Hofstadter's butterfly spectrum are of
fundamental interest to many research areas. 
Based upon slight modifications of existing cold-atom experiments, a
cold-atom realization of quantum maps with Hofstadter's butterfly
spectrum is proposed. Connections and differences between our
realization and the kicked Harper model are identified. This work
also exposes, for the first time, a simple connection between the
kicked Harper model and the kicked rotor model, the two paradigms of
classical and quantum chaos.

\end{abstract}
 \maketitle

The Harper model \cite{Harper} plays a fundamental role in many
research areas because it yields the famous Hofstadter's butterfly
spectrum \cite{Hfst,Lb}. This fractal spectrum, first discovered in
two-dimensional electron systems subject to a square lattice
potential and a perpendicular magnetic field \cite{Hfst}, has found
applications in studies of quantum Hall effect \cite{Albr,Hase},
the renormalization group \cite{Thouless}, high-temperature
superconductivity \cite{Mier}, to name a few. Various effects on
Hofstadter's butterfly spectrum were carefully examined
\cite{effects}. Systems with a butterfly spectrum should be also of
general interest to quantum phase transition studies because it
implies an infinite number of phases when some external parameters
are scanned. Systems with Hofstadter's butterfly spectrum were also
studied experimentally \cite{Albr,Kuhl}.

Hofstadter's butterfly can emerge in the quasi-energy spectrum of
periodically driven systems as well. In this context, the kicked
Harper model (KHM), adapted from the Harper model by considering a
delta-kicking potential, has attracted vast interests \cite{KHM}.
The Hamiltonian of the KHM is given by $H_{KHM}=(L/T)\cos(p)+K
\cos(q)\sum_n \delta (t-nT)$, where $L$ and $K$ are two system
parameters, $T$ is the kicking period, $p$ and $q$ ($q\in [0,2\pi]$)
are conjugate momentum and angle variables, with their commutation
relation defining the effective Planck constant $\hbar$, namely,
$[q,p]=i\hbar$ (in a flat phase space, to be more precise). The
Hilbert space with the periodic boundary condition in $q$ is spanned
by the eigenfunctions $|m\rangle$ of $p$, with $p|m\rangle=m\hbar
|m\rangle$, $\langle q|m\rangle=\exp(imq)/\sqrt{2\pi}$, and $m$
being an integer. The KHM quantum map associated with the unitary
evolution for each period $T$ is given by
\begin{eqnarray}
U_{KHM}=e^{-i\frac{L}{\hbar}\cos(p)}e^{-i\frac{K}{\hbar}\cos(q)}.
\label{KHmap}
\end{eqnarray}
Because the classical limit of the quantum map $U_{KHM}$ is chaotic
in general, this map has become a paradigm for understanding (i) how
a fractal quasi-energy spectrum affects the quantum dynamics and the
associated quantum-classical correspondence, and (ii) how the
underlying classical chaos affects the butterfly.  For experimental
realizations of the KHM, one early study proposed to use
Fermi-surface electrons in external fields \cite{Iomin}. Another
study showed that the system of a charged particle kicked by a
designed field sequence \cite{Dana} can be mapped onto the KHM.
However, these two proposals have not led to experiments.
Connections between the KHM and the so-called kicked harmonic
oscillator model were also noticed \cite{danaPRL}, but only for the
special case of $K=L$.

Using cold atoms periodically kicked by an optical lattice, about
ten laboratories worldwide
\cite{experiment1,experiment4,experiment2,experiment3} have realized
the so-called kicked-rotor model (KRM) \cite{casatibook} as another
quantum map paradigm. Using similar notation as above and in the
same Hilbert space as the KHM, the Hamiltonian of the KRM is given
by $H_{KRM} =p^{2}/2+ K\cos(q)\sum_n \delta (t-nT)$. Many variants
of the KRM, obtained by considering different types of kicking
sequences or additional external potentials, have also been
achieved. 
In these studies the experimental setup itself has also advanced,
from using thermal atoms to using a Bose-Einstein condensate (BEC)
\cite{experiment4,experiment2} that has very large coherence width.
These ongoing experimental efforts motivate the following bold and
important question: can these cold-atom laboratories working on the
KRM also realize the kicked Harper model or its variants by slightly
modifying their existing apparatus? If yes, quantum maps with
Hofstadter's butterfly spectrum can soon be experimentally realized
in many cold-atom laboratories, an entirely new generation of
experiments can be planned, and novel applications of cold-atom
researches may be established.

Stimulated by our early work seeking a potential connection between
KHM and a variant of KRM \cite{gongpre07}, a very positive answer to
the above question is indeed provided here. In particular, we show
that previous experimental setup for the so-called double-kicked
rotor model (DKRM) \cite{experiment3,DKR} already suffices for
synthesizing a quantum map displaying Hofstadter's butterfly,
provided that one quantum resonance condition therein is met and the
initial atom cloud is a BEC that has sufficient coherence width. The
butterfly spectrum associated with the obtained quantum map is
almost indistinguishable from the standard result previously
calculated for the KHM. We then show connections and dramatic
dynamical differences between our quantum map and the KHM. In
addition to experimental interests, the results should also motivate
more theoretical work on quantum maps with Hofstadter's butterfly
spectrum.

Consider then a DKRM that is already experimentally realized
\cite{experiment3}. Using the same notation as above, the
Hamiltonian of a DKRM can be written as $ H_{DKRM} = p^{2}/2+ K_{1}
\cos(q)\sum_{n}\delta (t-nT)+ K_{2}\cos(q)\sum_n \delta(t-nT-\eta)$.
Evidently, in addition to kicks at $t=nT$, the rotor in a DKRM is
also subject to kicks at $t=nT+\eta$. The associated quantum map
$U_{DKRM}$ for a period from $nT+0^{-}$ to $(n+1)T+{0}^{-}$ is given
by
 \begin{eqnarray}
 U_{DKRM}=e^{-i(T-\eta)\frac{p^2}{2\hbar}}e^{-i\frac{K_{2}}{\hbar}\cos(q)}
 e^{-i\eta\frac{p^2}{2\hbar}}e^{-i\frac{K_{1}}{\hbar}\cos(q)}.
 \end{eqnarray}
Remarkably, if we now require the parameter $T$ to satisfy the
quantum resonance condition of the KRM, i.e., $T\hbar=4\pi$, then
due to the discreteness of the momentum eigenvalues, one obtains
$e^{-iT\frac{p^2}{2\hbar}}=1$ when operating on any state in the
Hilbert space defined above. Under this resonance condition we are
able to reduce $U_{DKRM}$ to ${U}^{r}_{DKRM}$,
\begin{eqnarray}
 {U}_{DKRM}^{r}=e^{i\eta\frac{p^2}{2\hbar}}e^{-i\frac{K_{2}}{\hbar}\cos(q)}
 e^{-i\eta\frac{p^2}{2\hbar}}e^{-i\frac{K_{1}}{\hbar}\cos(q)}.
 \label{DKRM-r}
 \end{eqnarray} Because the cold atoms are
actually moving in a flat space rather than a compact angular space,
the quantum resonance condition is relevant only if the initial
quantum state is prepared in a definite quasi-momentum state. This
is certainly within reach of today's experiments. For example, two
recent experiments \cite{experiment2} studied directed transport in
a KRM on quantum resonance, with a delocalized BEC (with negligible
self-interaction) effectively realizing appropriate initial states
such as $|m=0\rangle$.  Note also
that the quantum map ${U}^{r}_{DKRM}$ offers a cold-atom realization
of a modified kicked-rotor model we recently proposed
\cite{gongpre07}, where the kinetic energy term can take ``negative"
values.

\begin{figure}
\begin{center}
\epsfig{file=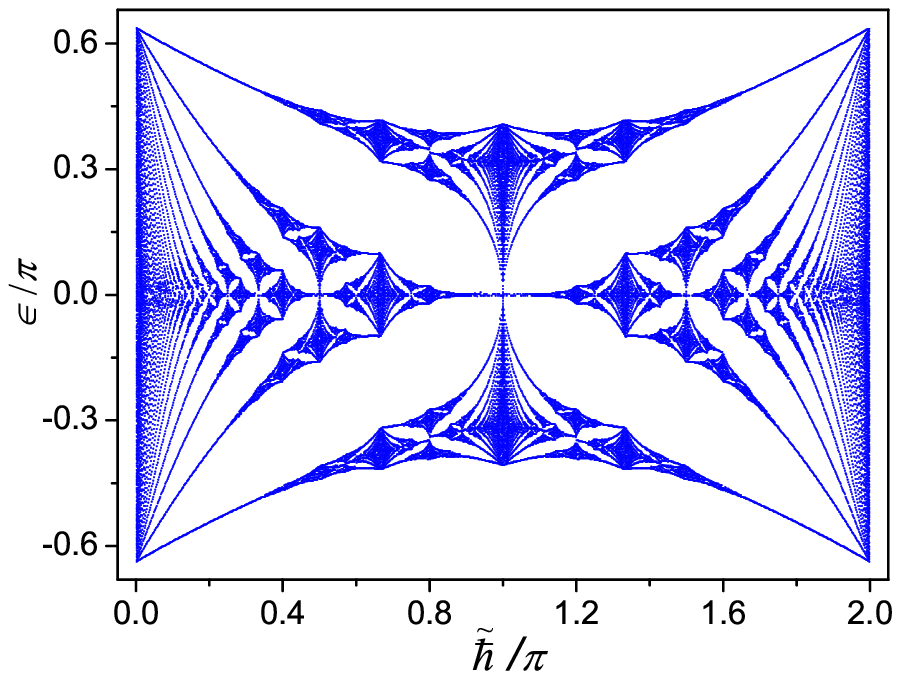,width=7.5cm}\vspace{-0.4cm}

\epsfig{file=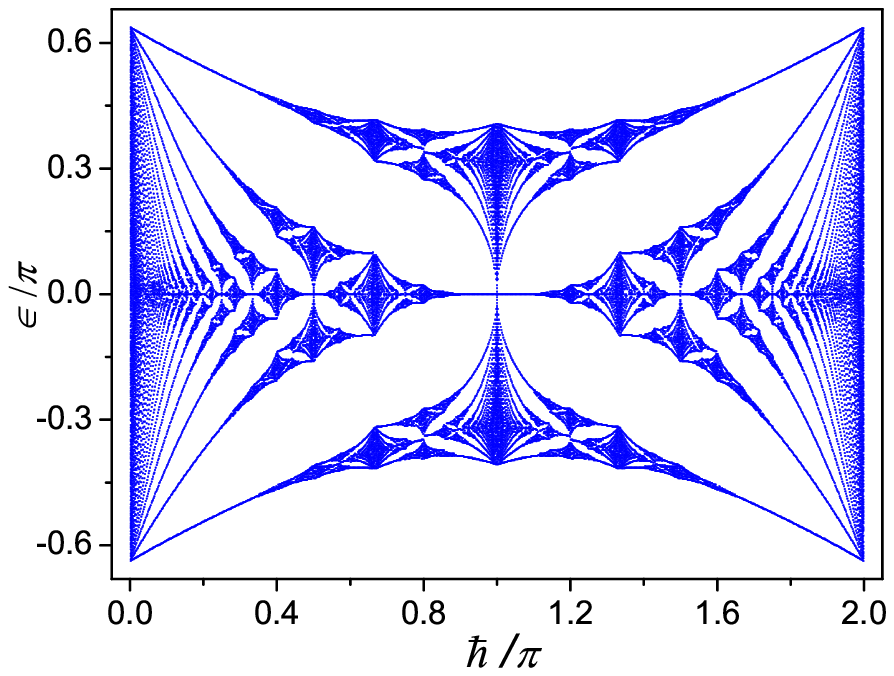,width=7.5cm}\vspace{-0.4cm} \caption{(Color
online) Quasi-energy spectrum (denoted $\epsilon$) of the quantum
map ${U}_{DKRM}^{r}$ given by Eq. (\ref{DKRM-r}) (top), compared
with that of the KHM map $U_{KHM}$ (bottom) given by Eq.
(\ref{KHmap}). $K_{1}/\hbar=K_{2}/\hbar=K/\hbar=L/\hbar=1$.
} \label{fig1}
\end{center}
\end{figure}

With the ${U}_{DKRM}^{r}$ realized above, we now present one key
numerical result of this work.  Figure \ref{fig1} displays the
calculated quasi-energy spectrum of ${U}_{DKRM}^{r}$ by the standard
diagonalization method \cite{KHM} as a function of
$\tilde{\hbar}\equiv\eta\hbar$, for $K_{1}/\hbar=K_{2}/\hbar=1$,
compared with that of $U_{KHM}$ as a function of $\hbar$, for
$K/\hbar=L/\hbar=1$. The map ${U}^{r}_{DKRM}$ is seen to generate a
beautiful Hofstadter's butterfly. Even more dramatically, the
butterfly of $U^{r}_{DKRM}$ resembles the previously calculated
butterfly of KHM \cite{KHM} to such a degree that the top panel
appears to be indistinguishable from the bottom panel in Fig.
\ref{fig1}.
The generalized fractal dimensions, denoted by $D_{q}$, have also
been calculated for many system parameters, confirming that the
spectrum is indeed a fractal in general [e.g.,  for $K_{1}=K_{2}=1$,
$\tilde\hbar=2\pi/(1+\sigma)$, $\sigma=(\sqrt{5}+1)/2$,
$D_{q=0}\approx 0.5$, same as the result for the KHM for
$\hbar=\tilde{\hbar}$, $K=L=1$].  Similar results have been found
for many other system parameters as well, so long as
$\hbar=\tilde{\hbar}$, $K_{1}=K$, and $K_{2}=L$.  To confirm a
fractal butterfly spectrum experimentally, one may connect the
associated characteristics of the quantum diffusion dynamics (e.g.,
time dependence of the survival probability, the diffusion exponent
etc.) with the spectrum \cite{fractal} or attempt to reconstruct the
spectrum by first reconstructing the time evolving wavefunction.

Results in Fig. \ref{fig1} suggest a strong connection between a
DKRM under quantum resonance and the KHM.  To uncover this
connection let us return to Eq. (\ref{DKRM-r}) and temporarily treat
it in a flat phase space without the periodic boundary condition.
Using the equality
$e^{i\eta\frac{p^{2}}{2\hbar}}f(q)e^{-i\eta\frac{p^{2}}{2\hbar}}=
f(q+\eta p)$, this treatment leads to
\begin{eqnarray}
{U}^{r}_{DKRM}\rightarrow \tilde{U}^{r}_{DKRM}
= e^{-i\frac{\tilde{K}_{2}}{\tilde{\hbar}} \cos(q+\tilde{p})}
e^{-i\frac{\tilde{K}_{1}}{\tilde{\hbar}}\cos(q)}, \label{KHform}
\end{eqnarray}
where $\tilde{p}\equiv \eta p$ is a rescaled momentum variable with
$\tilde{p}=-i\tilde{\hbar}\partial/\partial q$, $\tilde{K}_{1}=\eta
K_{1}$, and $\tilde{K}_{2}=\eta K_{2}$. Equation (\ref{KHform}) now
clearly resembles $U_{KHM}$ in Eq. (\ref{KHmap}), with the only
difference being that, the first exponential factor of
$\tilde{U}^{r}_{DKRM}$ in Eq. (\ref{KHform}) contains the
$\cos$-function of the angle plus the momentum, rather than just the
momentum.  This already partially rationalizes the strong
resemblance between the two panels  in Fig. \ref{fig1}.

\begin{figure}
\begin{center}
\epsfig{file=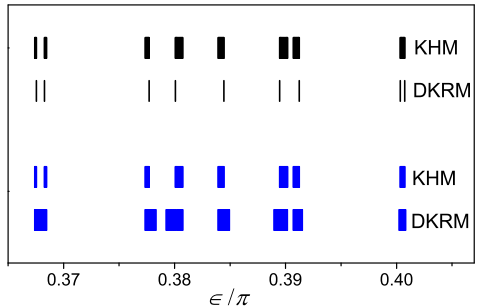,width=7.5cm}
\end{center}\vspace{-.8cm}
\caption{(Color online) Part of quasi-energy spectrum of
$U_{DKRM}^{r}$ in Eq. (\ref{DKRM-r}) and $U_{KHM}$ in Eq.
(\ref{KHmap}) for $\tilde{\hbar}=\hbar=\hbar_0=26\pi/41$ (the upper
two rows) and $\tilde{\hbar}=\hbar=2\pi-\hbar_0$ (the lower two
rows).
$\tilde{K}_{1}/\tilde{\hbar}=\tilde{K}_{2}/\tilde{\hbar}=K/\hbar=L/\hbar=1$.
The spectral differences between $U_{DKRM}^{r}$ and $U_{KHM}$ are
evident.} \label{fig2}
\end{figure}

So is the spectrum of $\tilde{U}^{r}_{DKRM}$ identical with that of
the KHM? Put differently, does there exist a unitary transformation
$\cal{G}$ to ensure
${\cal{G}}^{\dagger}(q+\tilde{p}){\cal{G}}=\tilde{p}$,
${\cal{G}}^{\dagger}q {\cal{G} }=q$? If such a ${\cal{G}}$ exists,
then ${\cal{G}}^{\dagger}\tilde{U}^{r}_{DKRM}{\cal{G}}$ becomes
precisely the $U_{KHM}$ in Eq. (\ref{KHmap}) (with
$\tilde{K}_{1}\rightarrow K$, $\tilde{K}_{2}\rightarrow L$, and
$\tilde{\hbar}\rightarrow \hbar$). 
Significantly, such a ${\cal{G}}$ does not exist for the Hilbert
space here. In particular, the above ${\cal{G}}$ transformation is
found to assume the analytical form
${\cal{G}}=e^{iq^{2}/2\tilde{\hbar}}$, which violates the periodic
boundary condition associated with $q\rightarrow q+2\pi$. As such,
the spectrum of $\tilde{U}^{r}_{DKRM}$, and hence also the spectrum
of ${U}^{r}_{DKRM}$ in (\ref{DKRM-r}), should contain substantial
elements that are absent in the KHM.

Indeed, as shown in Fig. 2, a more careful comparison does expose
spectral differences between $U_{DKRM}^{r}$ and $U_{KHM}$. Motivated
by this observation, we are also able to find some major differences
analytically. For example, for fixed $L/\hbar$ and $K/\hbar$, the
spectrum of $U_{KHM}$ is invariant upon a $\hbar$-change from
$\hbar_{0}$ to $2\pi-\hbar_{0}$. Such a symmetry does not exist in
the case of $U_{DKRM}^{r}$ (see Fig. 2).
A second example is for the special case of
$\hbar=\tilde{\hbar}=4\pi$. Therein the spectrum range can be easily
found, which is
$[-(\tilde{K}_{1}+\tilde{K}_{2})/\tilde{\hbar},(\tilde{K}_{1}+\tilde{K}_{2})/\tilde{\hbar}]$
for $U_{DKRM}^{r}$ and $[-(K+L)/\hbar,(K-L)/\hbar]$ for $U_{KHM}$
(if none of these range boundaries exceeds $\pm\pi$).

More insights emerge if we examine one interesting classical limit
of the DKRM, i.e., the $\tilde{\hbar}\rightarrow 0$ limit (by
letting $\eta\rightarrow 0$ with fixed $\tilde{K}_{1}$ and
$\tilde{K}_{2}$) while keeping $T\hbar=4\pi$. Denote
$(q_{l},\tilde{p}_{l})$ as a classical trajectory right before the
$l$th kick in the $\tilde{\hbar}\rightarrow 0$ limit of
$U^{r}_{DKRM}$. Then one obtains
$\tilde{p}_{2l+1}=\tilde{p}_{2l}+\tilde{K}_{2}\sin(q_{2l})$; $
q_{2l+1}=q_{2l}+\tilde{p}_{2l+1}$;
$\tilde{p}_{2l+2}=\tilde{p}_{2l+1}+\tilde{K}_{1}\sin(q_{2l+1})$; and
$q_{2l+2}=q_{2l+1}-\tilde{p}_{2l+2}$.  We stress that this classical
limit is obtained under quantum resonance, and is hence unrelated to
the direct classical analog \cite{casatipre} of the quantum DKRM.
Upon making a classical canonical transformation
$(q,\tilde{p}+q)\rightarrow (Q,\tilde{P})$, we obtain
$\tilde{P}_{2l+2}=\tilde{P}_{2l}+\tilde{K}_{2} \sin(Q_{2l})$;
$Q_{2l+2}=Q_{2l}-\tilde{K}_{1} \sin(\tilde{P}_{2l+2})$, which is
precisely the classical map of the kicked Harper model. This finding
hence firmly binds the butterfly spectrum of $U_{DKRM}^{r}$ with the
standard KHM. The emergence of the classical KHM map from the
$\tilde{\hbar}\rightarrow 0$ limit of $U_{DKRM}^{r}$ further
demonstrates that the spectral differences between ${U}^{r}_{DKRM}$
and $U_{KHM}$ arise from genuine quantization effects.  Indeed, it
is the periodic boundary condition in the quantization that
disallows the above-mentioned unitary transformation ${\cal G}$ as
the quantum analogy of the classical canonical transformation
$(q,\tilde{p}+q)\rightarrow (Q,\tilde{P})$.

\begin{figure}
\vspace{-.1cm}
\begin{center}
\epsfig{file=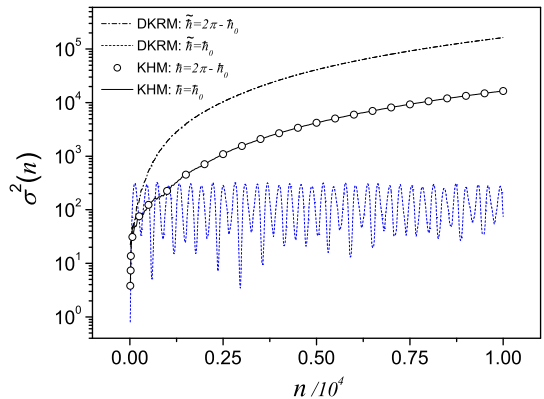,width=8.cm}\vspace{-.8cm}
\end{center}

\caption{(Color online) Dynamics of $m$-variance (denoted
$\sigma^{2}$) for quantum maps $U_{DKRM}^{r}$ and $U_{KHM}$, for
the initial state $|0\rangle$. The values of $\tilde{\hbar}=\hbar$
are given by $\hbar_0$ or $2\pi-\hbar_0$, with $\hbar_0=26\pi/41$.
$\tilde{K}_{1}/\tilde{\hbar}=\tilde{K}_{2}/\tilde{\hbar}=L/\hbar=K/\hbar=1$.
The map $U_{KHM}$, not $U_{DKRM}^{r}$, is seen to be invariant
upon the change $\hbar_{0}\rightarrow 2\pi-\hbar_{0}$.  Note also
that one case associated with $U_{DKRM}^{r}$ displays
localization, and all the others show quadratic diffusion. }
\label{fig3}
\end{figure}

The spectral differences between ${U}^{r}_{DKRM}$ and $U_{KHM}$ are
found to result in profound consequences in the quantum dynamics.
One excellent example is shown in Fig. \ref{fig3}, demonstrating
clearly that the butterfly associated with ${U}^{r}_{DKRM}$ violates
a symmetry property of the butterfly associated with $U_{KHM}$. Note
also that in one case of ${U}^{r}_{DKRM}$ shown in Fig. \ref{fig3},
the quantum diffusion displays evident localization that is in clear
contrast to the KHM dynamics. This localization behavior suggests
that the width of the sub-bands of the butterfly is effectively
zero, in agreement with the result shown as the second row in Fig.
2. Indeed, in our numerical analysis the associated band width is
found to be less than $10^{-16}$.

To further motivate interests in the new quantum map
${U}^{r}_{DKRM}$ we also present in Fig. 4 the time-dependence of
the momentum distribution profile, for a more generic
$\hbar=\tilde{\hbar}$ that is irrational with $\pi$. Results therein
show again striking differences between ${U}^{r}_{DKRM}$ and
$U_{KHM}$, especially in that the former case shows a staircase
structure in the profile. Unlike previous observations of analogous
staircase structure with a classical origin in an off-resonance DKRM
\cite{DKR}, the staircase structure here is purely quantum
mechanical. Indeed, both the localization shown in Fig. \ref{fig3}
and the staircase profile shown in Fig. 4 can be related to the
unique blocked band structure of ${U}^{r}_{DKRM}$ in the momentum
representation \cite{note1}.  The important lesson here is that many
important features of a quantum map can be hidden in the overall
pattern of its butterfly spectrum. For experimental interests, we
note that one may tune the value of $\tilde{\hbar}$ and other system
parameters to generate different block sizes of ${U}^{r}_{DKRM}$
\cite{note1}, thus attaining staircase steps of less height and
hence more accessible to experiments. For example, for
$K_1=K_2=14.4$, $\tilde\hbar\approx 118 \pi/61$, the associated
staircase steps in $P(m)$ have a width of $61$, and a height $2-3$
orders of magnitude smaller.

\begin{figure}
\begin{center}
\epsfig{file=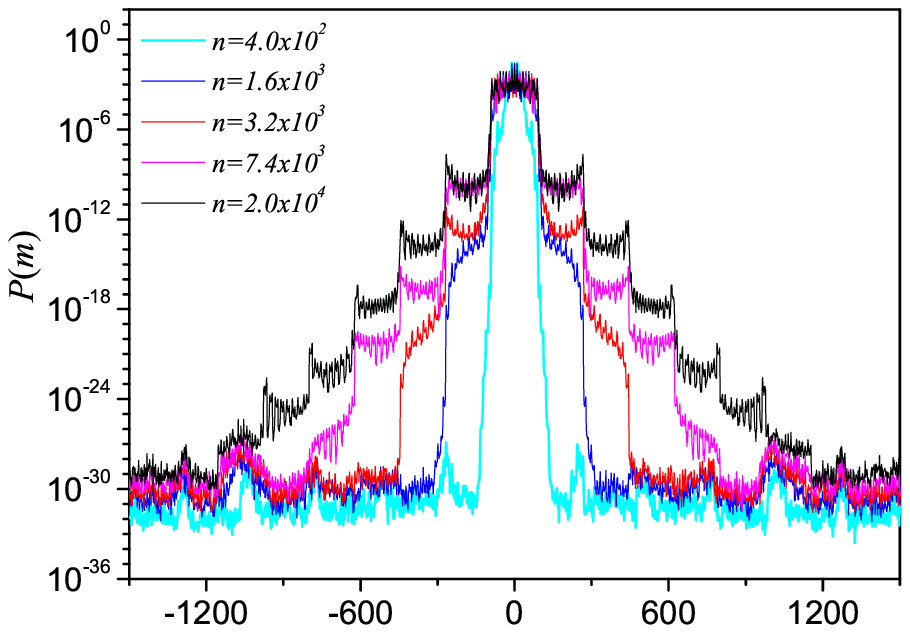,width=7.8cm}\vspace{-.4cm}

\epsfig{file=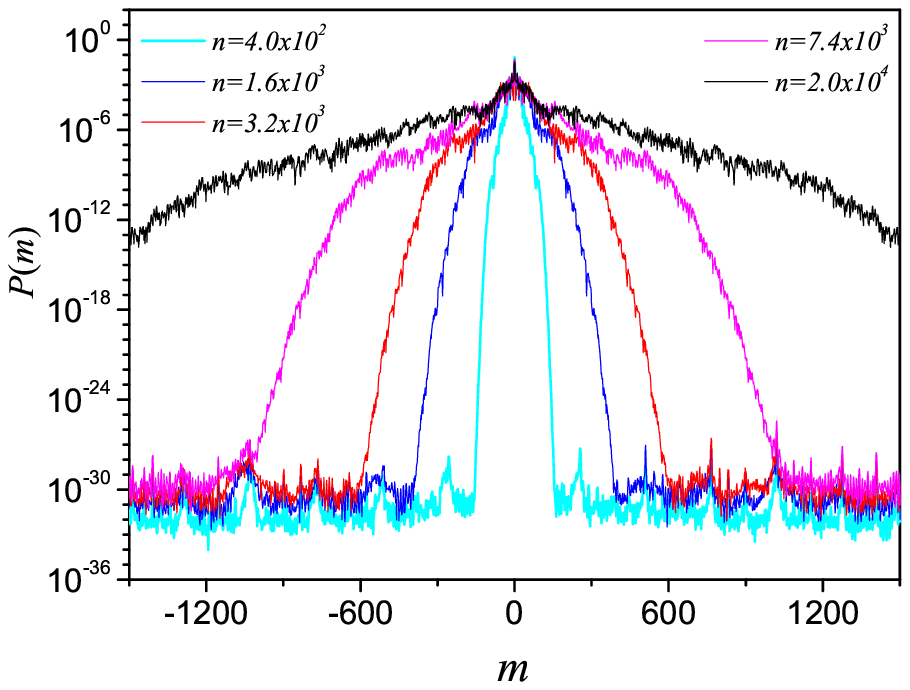,width=7.8cm}\vspace{-.6cm}\end{center}
\caption{(Color online) Momentum distribution profile $P(m)$ vs $m$,
evolving from the initial state $|0\rangle$, for  $U_{DKRM}^{r}$
(top) and $U_{KHM}$ (bottom), for $\tilde{\hbar}=\hbar=2$,
$\tilde{K}_{1}=\tilde{K}_{2}=K=L=3.7$. The staircase structure seen
for $U_{DKRM}^{r}$ does not exist for $U_{KHM}$, despite the fact
that their two butterfly spectrum is hardly distinguishable.}
\label{fig5}
\end{figure}

Cold-atom realizations of the non-kicked Harper model using static
optical lattices were proposed before \cite{holthaus}. However, due
to the deep lattice approximation therein they cannot be extended
for the kicked Harper model. Based on already available experimental
techniques that can achieve a double-kicked rotor model tuned on
quantum resonance, here we have proposed a rather simple cold-atom
realization of a variant of the kicked Harper model.
The results should open up a new generation of cold-atom experiments
on quantum maps with a butterfly spectrum. This work also
establishes, for the first time, a direct connection between the
kicked-rotor model and the kicked Harper model, arguably the two
most important paradigms of classical and quantum chaos.

{\bf Acknowledgments}: Very constructive comments made by Dr.
Zai-Qiao Bai are gratefully acknowledged. We also thank Prof. C.-H.
Lai for his support and encouragement. J.W. acknowledges support
from Defence Science and Technology Agency (DSTA) of Singapore under
agreement of POD0613356. J.G. is supported by the start-up funding
(WBS grant No. R-144-050-193-101 and No. R-144-050-193-133) and the
NUS ``YIA'' funding (WBS grant No. R-144-000-195-123), National
University of Singapore.


\begin{thebibliography}{99}
\bibitem{note1} For small $K_{1}/\hbar=K_{2}/\hbar$, we analytically
find $\langle m|U^{r}_{DKRM}|m+1\rangle \propto
\{1+\exp[-i\tilde{\hbar}(2m+1)/2\}$ \cite{gongpre07}, which can be
zero. Hence, to a good approximation this matrix is blocked,
especially for $\tilde{\hbar}=2\pi(2k+1)/(2n+1)$, $k$, $n$ being
integers.
\end{thebibliography}
\end{document}